\documentclass[lettersize,journal]{IEEEtran}
\usepackage{amsmath,amsfonts} 
\usepackage{algorithmic}
\usepackage{algorithm}
\usepackage{array,subfigure,ltablex,tabularx,setspace}
\usepackage{textcomp}
\usepackage{stfloats}
\usepackage{url}
\usepackage{verbatim}
\usepackage{graphicx} 
\usepackage{cite}
\usepackage{xcolor}
\usepackage{mathtools}
\usepackage{environ}
\usepackage{pifont}

\usepackage{orcidlink}  

\renewcommand{\alpha}{\kappa}
\renewcommand{\alpha}{\kappa}

\begin{document}

\title{Accurate Closed-Form Solution for Symbol Error Probability in Hexagonal QAM}
\author{Sukhsagar,~\IEEEmembership{Student Member,~IEEE},
\thanks{Sukhsagar is the Discipline of Centre for Advanced Electronics, Indian Institute of Technology Indore, Indore 453552, India (e-mail: phd2301191002@iiti.ac.in).}
\thanks{Vimal Bhatia is the Discipline of Electrical Engineering, Indian Institute of Technology Indore, Indore 453552, India (e-mail:  vbhatia@iiti.ac.in).}
Nagendra Kumar,~\IEEEmembership{ Member,~IEEE},
\thanks{Nagendra Kumar is with the Department of Electronics and Communication Engineering, National Institute of Technology, Jamshedpur 831014, India (e-mail: kumar.nagendra86@gmail.com).} 
Vimal Bhatia,~\IEEEmembership{ Senior Member,~IEEE}}




\maketitle

\begin{abstract}
Future communication systems are anticipated to facilitate applications requiring high data transmission rates while maintaining energy efficiency. Hexagonal quadrature amplitude modulation (HQAM) offers this owing to its compact symbol arrangement within the two-dimensional (2D) plane. Building on the limitations of the current approaches, this letter presents a straightforward and precise approximation for calculating the symbol error probability (SEP) of HQAM in  additive white Gaussian noise (AWGN) channel. The analytical and simulation results align across several signal-to-noise ratios (SNR). In addition, the proposed approximation is effective for accurately estimating the SEP of HQAM in scenarios with slow-fading, including those subject to Rayleigh fading.
\end{abstract}


\section{Introduction}
\IEEEPARstart{W}{ith} the global deployment of 5G and research on 6G focusing on the high-data-rate multimedia applications with minimal energy consumption. The same can be achieved by employing optimized bandwidth and power-efficient strategies. Energy-efficient high-speed communication network in the future wireless systems involves minimizing the average transmit power while ensuring a certain bit error rate (BER) or symbol error rate (SER). In order to maintain SER performance at higher data rates within the available bandwidth, it typically becomes necessary to boost transmit power.
This need has led recent research to focus on higher order 2D constellations, which offer high data rates in an energy-efficient manner, as well as more power-efficient 2D hexagonal-shaped constellations, known as HQAM \cite{singya2021survey}. In a 2D signal constellation, the SER is primarily influenced by the minimum distance between neighboring constellation points and the average symbol energy, which relates to the mean squared distance of the points from the origin.
Taking these factors into account, an optimal 2D hexagonal lattice-based HQAM constellation is suggested. In this setup, if the minimum distance between two adjacent points is \(2d\), the area of the hexagonal region is \(2\sqrt{3}d^2\). This area is 0.866 times that of a rectangular region with the same dimensions, resulting in an approximate gain of 0.6 dB over rectangular regions. HQAM provides the most compact 2D packing \cite{conway2013sphere}, which lowers both the peak and average power of the constellation, thereby improving its power efficiency compared to other constellations. HQAM shows exceptional performance at high SNR, offering an asymptotic SNR improvement of 0.825 dB over square QAM (SQAM) \cite{gallager1984efficient}. A simplified maximum likelihood (ML) detection method has been proposed for a specific type of HQAM called triangular QAM (TQAM) \cite{park2007triangular}, as the arrangement and spacing of the constellation points on the 2D plane are the same for both HQAM and TQAM.
It is crucial to recognize that the symbols in an HQAM constellation are placed at the centers of equilateral hexagons \cite{simon1973hexagonal} and in a TQAM constellation, the symbols are positioned at the corners of equilateral triangles \cite{singya2021survey}. HQAM constellations are divided into regular HQAM (R-HQAM) and irregular HQAM (I-HQAM) depending on their arrangement relative to the origin. R-HQAM exhibits symmetry around the origin, whereas I-HQAM does not adhere to this symmetry, leading to circular patterns as the modulation order (M) rises. However, I-HQAM constellations tend to be more compact than R-HQAM.
A key metric for assessing the performance of a modulation scheme is the SEP. However, the hexagonal lattice structure of HQAM makes it challenging to calculate the SEP precisely. The SEP approximation for TQAM in AWGN presented in \cite{park2012performance} employs a nearest-neighbor (NN) method that performs well primarily under high SNR conditions. Although the SEP evaluation of TQAM in \cite{qureshi2016sep} offers greater accuracy compared to \cite{park2012performance}, it is still not exact, contrary to the assertions made in \cite{qureshi2016sep}.
Additionally, the SEP analyses in \cite{simon1973hexagonal} involve numerical integration approaches that result in complex expressions.  In \cite{rugini2016symbol}, the SEP approximation for HQAM employed the average number of nearest neighbors (NNs) and incorporated an adjustment factor to enhance the approach from \cite{park2012performance}.
In \cite{sadhwani2018simple}, a method for approximating the SEP of HQAM is derived, but compared to \cite{rugini2016symbol}, the relative error and absolute error (AE) are higher, leading to a decrease in accuracy. 
Furthermore, in \cite{oikonomou2022error}, an upper bound and a mathematical derivation for the SEP of HQAM are proposed, along with the derivation of a low-complexity detection scheme.
While it aimed to reduce complexity compared to \cite{rugini2016symbol} and \cite{sadhwani2018simple}, the AE is high, thereby resulting in decreased accuracy. 

This letter introduces a precise and straightforward analytical expression for the SEP with reduced computational effort. Moreover, we validate the proposed approach for Rayleigh faded channels, highlighting its capability for simple yet accurate analysis of HQAM in slow-fading channels.

\section{ System Model }
\begin{figure}[!t]
\centering
\includegraphics[width=7cm,height=6cm]{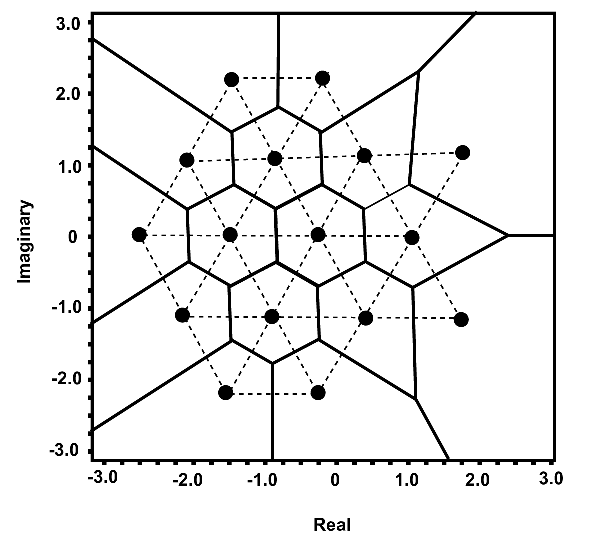}
\caption{16 HQAM Constellation.}
\label{fig_1}
\end{figure}

An M-ary HQAM constellation, depict ed in Fig. \ref{fig_1}, is characterized by a set $A_{M} = \{{s_{i}\in{R^2}}, i = 0,..., M-1\}$ containing ${M}$ symbols, where ${A_{M}}$ is a subset of the infinite grid $\ {\mathcal{G}} = \{{v\in{R^2}}:v =n_1v_1+n_2v_2+z_0\}$ with $v_1 = [d/2,\sqrt{3}d/2]^T$ and $v_2 = [d/2,-\sqrt{3}d/2]^T$ as the basis vectors of the 2D grid. Here, \(d > 0\) denotes the smallest distance between points, \(n_1\) and \(n_2\) are integer coefficients, and \(z_0\) is an offset in the 2D plane. 
The ML decision regions, illustrated in Fig. \ref{fig_1} feature central areas that form equilateral hexagons with internal angles of \(2\pi/3\). The outer regions may include one or more angles of \(2\pi/3\).


\begin{figure}[!t]
\centering
\includegraphics[width=4cm,height=4cm]{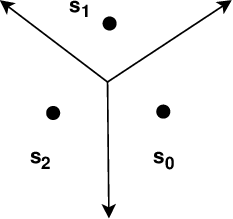}
\caption{3-PSK Constellation.}
\label{fig_2}
\end{figure}
Moreover, we can describe the received signal at the destination node as
 \begin{equation}
\label{eqn_1}
r =  s + n_0,
\end{equation}
where \(s\) represents the transmitted symbol and \(n_0\) denotes AWGN at the receiver with zero mean and variance \(N_0/2\), where \(N_0\) is the noise spectral density.

\section{Exact SEP in AWGN Channels}
An estimated formula for the SEP in an M-ary signal set within an AWGN channel was developed in \cite{park2012performance} and can be expressed as
\begin{equation}
\label{eqn_2}
  P_e = A Q\left(\frac{d}{2\sigma}\right),
\end{equation}
where \( Q(\cdot) \) represents the Gaussian \( Q \)-function, defined as \( Q(z) = \frac{1}{\sqrt{2\pi}} \int_{z}^{\infty} e^{-\frac{t^2}{2}} \, dt \), \( d = \frac{\sqrt{\alpha \gamma_s}}{2\sigma} \) represents the NNs distance, \( \sigma \) is the standard deviation of the AWGN, \( \gamma_s \) denotes the average SNR, and \( \alpha \) is specified in \cite[Table 6]{singya2021survey} for various constellation orders. Moreover, \( A \) represents the average number of NNs, calculated as \( A = \frac{1}{M} \sum_{i=0}^{M-1} A(i) \), where \( A(i) \) is the count of NNs for the symbol \( s_i \).
\begin{table}
\begin{center}
\caption{parameters B for the R-HQAM and I-HQAM .}
\label{tab_1}
\begin{tabular}{| c | c | c |}
\hline
Order & Regular & Irregular\\
M & HQAM &   HQAM \\
\hline
4& $1.9977$ & $1.9977$\\
\hline
8&$3.4960$ & $3.4960$\\ 
\hline
16& $4.4948$, & $4.4948$\\
\hline
32& $5.4937$, & $5.4937$\\
\hline
64& $6.1180$ & $6.2428$\\
\hline
128&$6.6174$& $6.7422$\\
\hline 
256& $7.0232$ & $7.1168$\\
\hline
512& $7.3080$ & $7.3704$\\
\hline
1024& $7.4992$ & $7.5382$\\
\hline
2048& $7.6416$ & $7.7168$\\
\hline
\end{tabular}
\end{center}
\end{table}

\begin{table}
\begin{center}
\caption{AE of 256-HQAM.}
\label{tab_2}
\begin{tabular}{| c | c | c |c |c|}
\hline
SNR & AE&AE&AE &AE \\
(dB)& App. (5)& [8] & [9] & [10]\\
\hline
-4& 0.0039&0.0041& 0.1306 &  0.0046\\
\hline
-1&   0.0045& 0.0056& 0.1305 &0.0050\\
\hline
  2&0.0048& 0.0072& 0.1287&  0.0052\\
  \hline
 5&0.0051& 0.0095&0.1237 & 0.0058\\
 \hline
 8&   0.0062& 0.0132&0.1127&0.0077\\
 \hline
 11& 0.0049& 0.0152&0.0872&0.0081\\
 \hline
 14& 0.0053& 0.0192&0.0460&0.0110\\
 \hline
 17& 0.0059&0.0221 &0.0085&0.0135\\
 \hline
20 &0.0088&0.0236&0.0458&0.0150\\
\hline
23 &0.0086&0.0169&0.0357&0.0093\\
\hline 
\end{tabular}
\end{center}
\end{table}
As an illustration, the simplest form of HQAM, specifically the 3-PSK is depicted in Fig. \ref{fig_2}. The transmitted symbol \( s_0 \) is surrounded by two NNs, \( s_1 \) and \( s_2 \). Given that \( s_1 \) is also a NN of \( s_2 \) (and vice versa), \( s_1 \) and \( s_2 \) are considered a pair of adjacent NN to \( s_0 \). The error region in the NN approximation is covered by two overlapping half-planes, leading to an overestimation of the SEP. This is illustrated in Fig. \ref{fig_2}. Therefore, the precise expression for the SEP in 3-PSK is provided as
\begin{equation}
\label{eqn_3}
  P_{3-PSK} = 2Q\left(\sqrt{\alpha\gamma_s}\right)-C, 
\end{equation}
where $C$ represents the correction factor \cite{rugini2016symbol}, addressing the issue of double counting due to the overlap. Given that 3-PSK is a simplex signal \cite{proakis2008digital}, $C$ can be expressed as
\begin{eqnarray}
\label{eqn_4}
   C = &\frac{1}{\sqrt{2\pi}} \int_{-\infty}^{\infty}\left[2Q(\sqrt{\alpha\gamma_s})-\{1-(1-Q(z))^2\}\right]\nonumber\\&
   \times e^{-{(z-{\sqrt{2\alpha \gamma_s}})}^2/2}\,dz.
   \end{eqnarray}

\textbf{Theorem 1:} The novel accurate closed-form expression of SEP for HQAM is given as
\begin{align}
\label{eqn_5}
  P_{HQAM} = AQ\left(\sqrt{\alpha\gamma_s}\right)-B Q\left(\sqrt{\frac{10}{11}\alpha\gamma_s}\right)Q\left(\sqrt{\frac{1}{3}\alpha\gamma_s}\right),
\end{align}
\textbf{\textit{Proof:}}
In Fig. \ref{fig_1}, several overlaps occur at angles of \(2\pi/3\), each associated with a pair of neighboring NNs. 
Furthermore, by applying the correction method, we can extend (\ref{eqn_3}) to any HQAM constellation as
\begin{align}
\label{eqn_6}
  P_{HQAM} = AQ\left(\sqrt{\alpha \gamma_s}\right)-A_c C,
\end{align}
where corresponding values of $A$ and $A_{c}$ are given in \cite[Table 6]{singya2021survey}. Further, by substituting the value of $C$ from (\ref{eqn_10}), (\ref{eqn_12}), (\ref{eqn_14}), and (\ref{eqn_16}), we can simplify (\ref{eqn_11}) to the form shown in (\ref{eqn_5}), where $B=1.3318 A_c$.

Moreover, we evaluate numerical values of the parameter $B$ for several constellations, as detailed in Table \ref{tab_1}.
The approximations presented in (\ref{eqn_5}) are closed-form solutions that have not been previously documented in the literature. It is noteworthy that these approximations offer lower complexity for a given SNR compared to those in \cite{rugini2016symbol}, \cite{sadhwani2018simple}, and \cite{oikonomou2022error}. Consequently, for a specified SNR \(\gamma_s\), computing the SEP using (5) requires just 5 multiplications and 2 function evaluations of \(Q(\sqrt{.})\). In addition, as shown in Table \ref{tab_2}, the AE is lower as compared to approximations in \cite{rugini2016symbol}, \cite{sadhwani2018simple}, and \cite{oikonomou2022error}. Therefore, the proposed approximation demonstrates significantly higher accuracy than the existing approximations with lower computational complexity.

\section{SEP for HQAM Under Rayleigh Distributed Environment}
Under frequency-flat Rayleigh fading conditions, \( E_s = \theta^2 \bar{E}_{s} \) denotes energy per symbol with a random amplitude \( \theta \) and \( \bar{E}_{s} \) is the average symbol energy. SEP for an HQAM modulation scheme in a Rayleigh fading channel can be mathematically expressed using a well-known probability density function (PDF) based approach as follows:

\begin{equation}
\label{eqn_7}
P_{e}^{\text{R}} = \int_0^\infty P_{\text{AWGN}}(\theta) f_\Theta(\theta) \, d\theta,
\end{equation}
where \( P_{\text{AWGN}}(\theta) \) is conditional SEP and $f_\Theta(\theta)$ denotes PDF for the random amplitude \( \theta \), is given in \cite{rugini2016symbol} as:
\begin{equation}
\label{eqn_8}
f_\Theta(\theta) = 2\theta e^{-\theta^2}, \quad \theta \geq 0.
\end{equation}
Next, we simplify $P_{AWGN}(\theta)$, as given in (\ref{eqn_7}), using $P_{HQAM}$ from 
(\ref{eqn_5}) calculated with \( \gamma_s = \theta^2 \bar{\gamma}_{s} \), where  \( \bar{\gamma_s} =  \frac{\bar{E_s}}{N_0} \) represents the average SNR.

By substituting (\ref{eqn_5}) and (\ref{eqn_8}) into (\ref{eqn_7}) and solving the integrals with respect to the variable $\bar{\gamma_s}$ using \cite[(5) and (7)]{beaulieu2006useful}, an expression for the average SEP for general-order HQAM can be derived as:
\begin{align}
\label{eqn_9}
P_{e}^{\text{R}} &=  \frac{A}{2} \left( 1 - \sqrt{\frac{\frac{1}{2}  \bar{\gamma_s}  \alpha}
{1 + \frac{1}{2}  \bar{\gamma_s}  \alpha}} \right) - B \Bigg[\frac{1}{4} - \frac{1}{2\pi}\Bigg\{\sqrt{\frac{5 \bar{\gamma_s} \alpha}{11 + 5 \bar{\gamma_s} \alpha}} \nonumber\\&
\times \arctan \left(\sqrt{\frac{330 + 150 \bar{\gamma_s} \alpha}{55 \bar{\gamma_s} \alpha}}\right)+\sqrt{\frac{\bar{\gamma_s} \alpha}{6 + \bar{\gamma_s} \alpha}}\nonumber\\&
\times \arctan \left(\sqrt{\frac{66 + 11 \bar{\gamma_s} \alpha}{30 \bar{\gamma_s} \alpha}}\right)\Bigg\}\Bigg].
\end{align}

\section{Simulation Results}
\begin{figure}[!t]
\centering
\includegraphics[width=8cm,height=4.5cm]{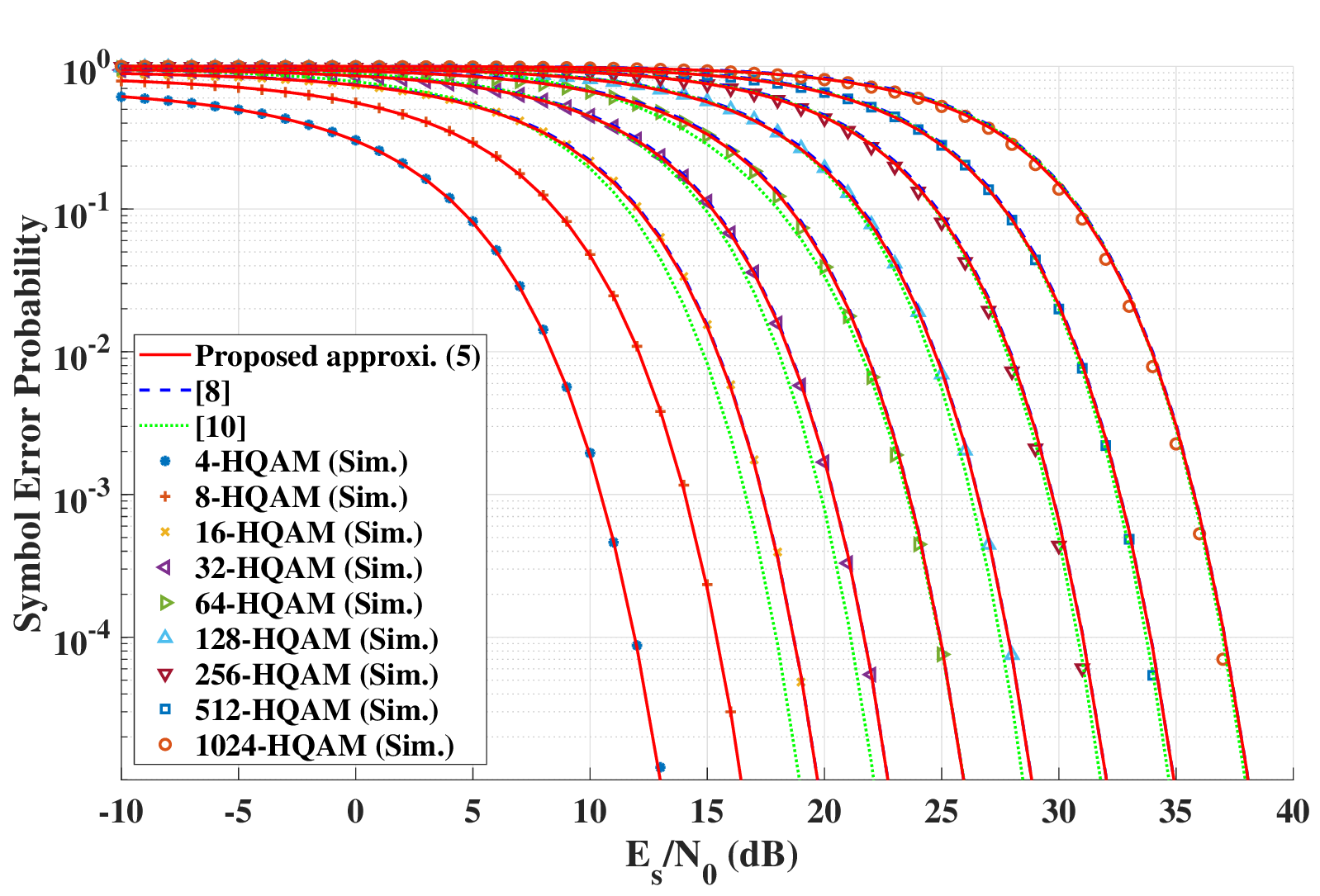}
\caption{Simulated and theoretical SEP of HQAM Constellations.}
\label{fig_3}
\end{figure}

In Fig. \ref{fig_3}, the approximated SEP derived from (\ref{eqn_5}) for regular HQAM constellations is presented and validated through simulation results against $ \frac{E_s}{N_0} = \frac{E_b}{N_0} \log_2 M $. The figure shows that the approximated and simulated curves align well, confirming accuracy of the proposed expression across both low and high SNR values for all orders of HQAM constellations.
Additionally, the proposed approximation has been compared with the existing results provided in \cite{rugini2016symbol} and \cite{oikonomou2022error}. The comparison confirms that the proposed results demonstrate superior SEP performance and exhibit greater accuracy than those presented in both \cite{rugini2016symbol} and \cite{oikonomou2022error} across most of the SNR values.

\begin{figure}[!t]
\centering
\includegraphics[width=8cm,height=4.5cm]{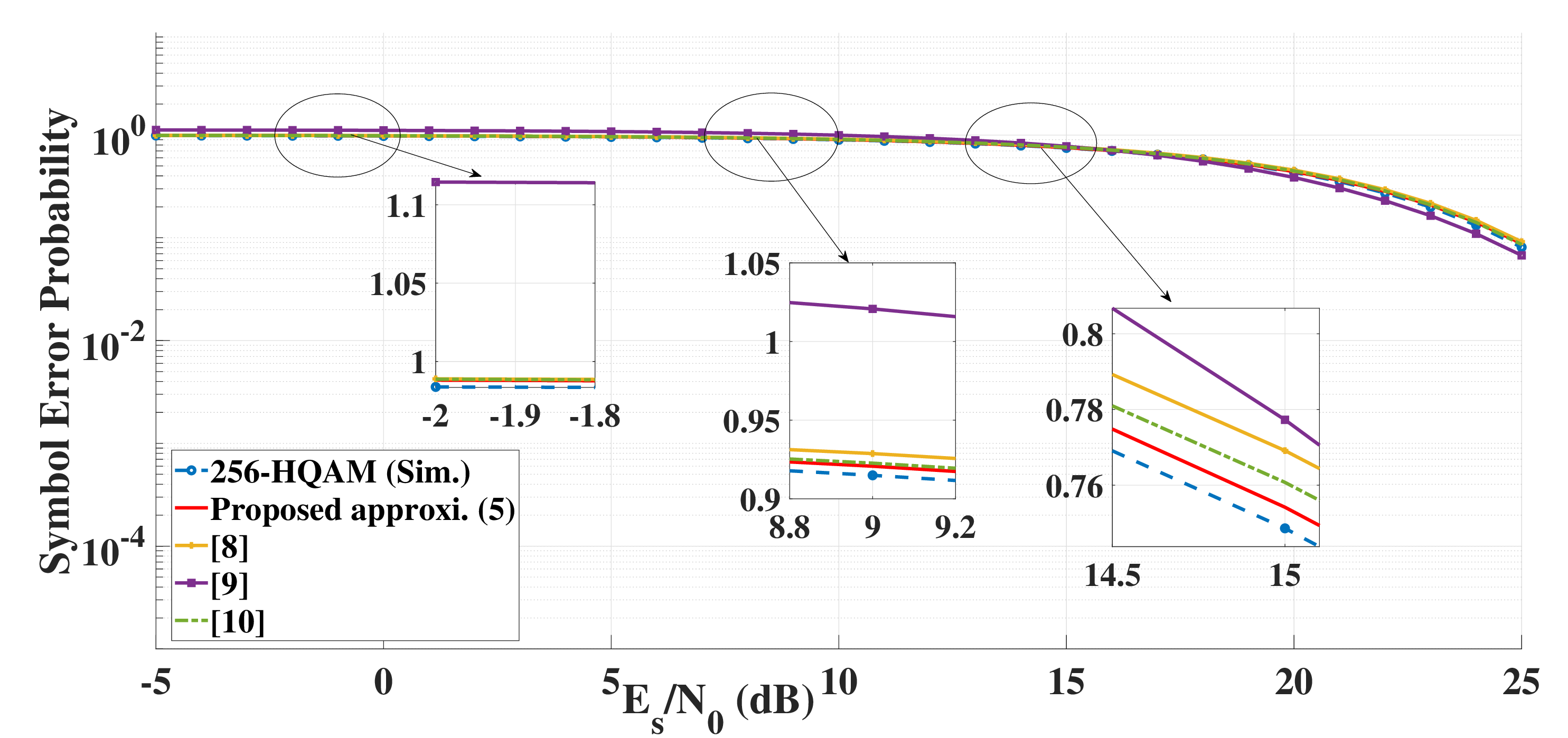}
\caption{Compares the SEP approximation v/s received SNR.}
\label{fig_4}
\end{figure}

In Fig. \ref{fig_4}, the SEP approximation derived from (\ref{eqn_5}) is compared with the simulated results and the previously established approximations from \cite{rugini2016symbol}, \cite{sadhwani2018simple}, and \cite{oikonomou2022error} for higher-order 256-HQAM. The simulated results aligns more closely with the proposed approximation than all existing approximations, further validating accuracy of the proposed method.

\begin{figure}[!t]
\centering
\includegraphics[width=8cm,height=4.5cm]{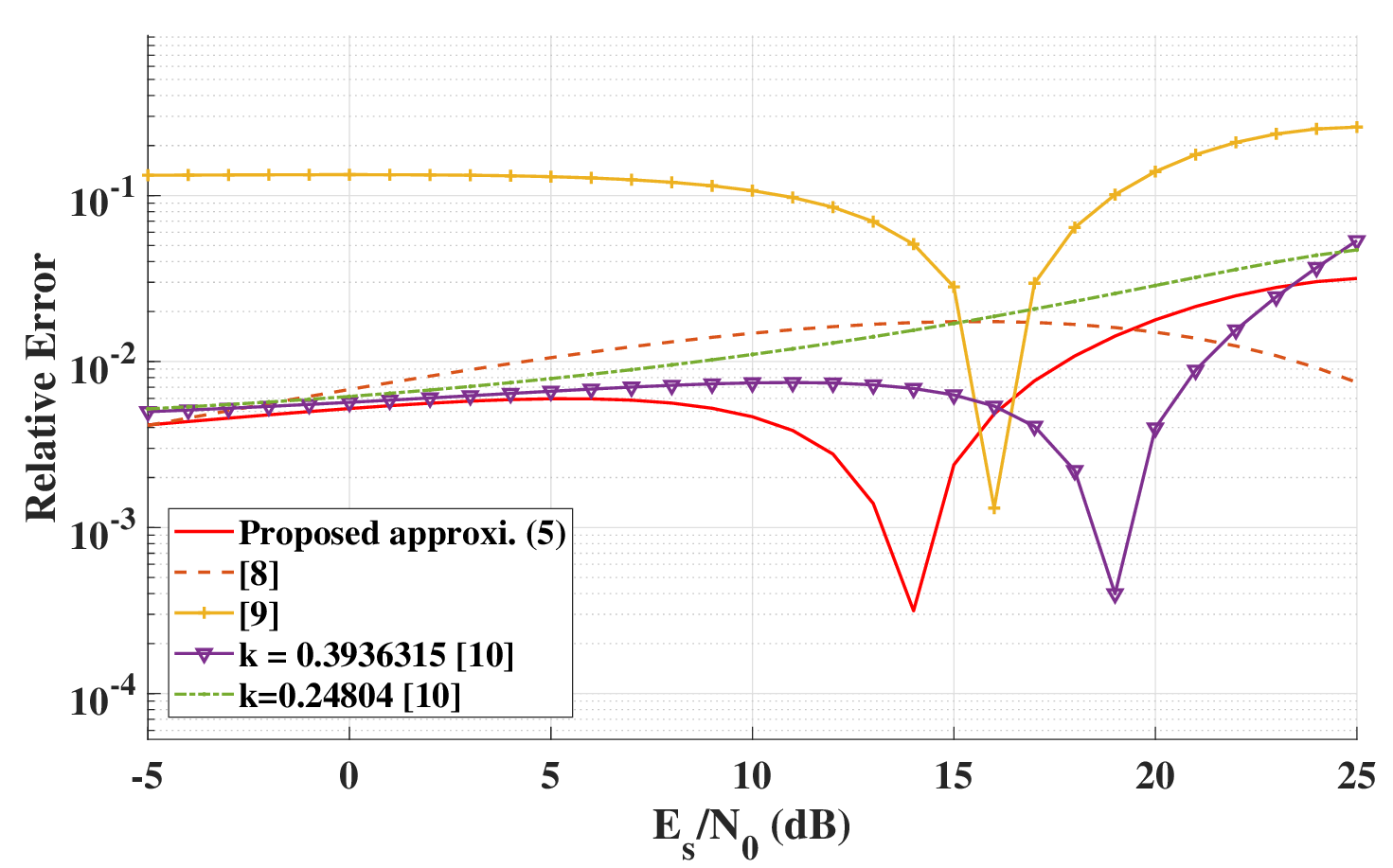}
\caption{Compares the relative error v/s received SNR of 256-HQAM.}
\label{fig_5}
\end{figure}

Fig. \ref{fig_5} shows the relative error in the exact SEP for the 256-HQAM constellation using the proposed approximation as compared to the methods in \cite{rugini2016symbol}, \cite{sadhwani2018simple}, and \cite{oikonomou2022error}. The relative error, defined as \( \beta = \frac{|\text{approximated SEP} - \text{exact SEP}|}{\text{exact SEP}} \), quantifies how much the approximation deviates from the exact SEP value. It is evident from Fig. \ref{fig_5} that the proposed method outperforms those in \cite{rugini2016symbol}, \cite{sadhwani2018simple}, and \cite{oikonomou2022error}. Additionally, the proposed approximation is highly accurate, with the relative error remaining below \( 10^{-2} \) from -5dB to 18dB, which is less than 1\%.

\begin{figure}[!t]
\centering
\includegraphics[width=8cm,height=4.5cm]{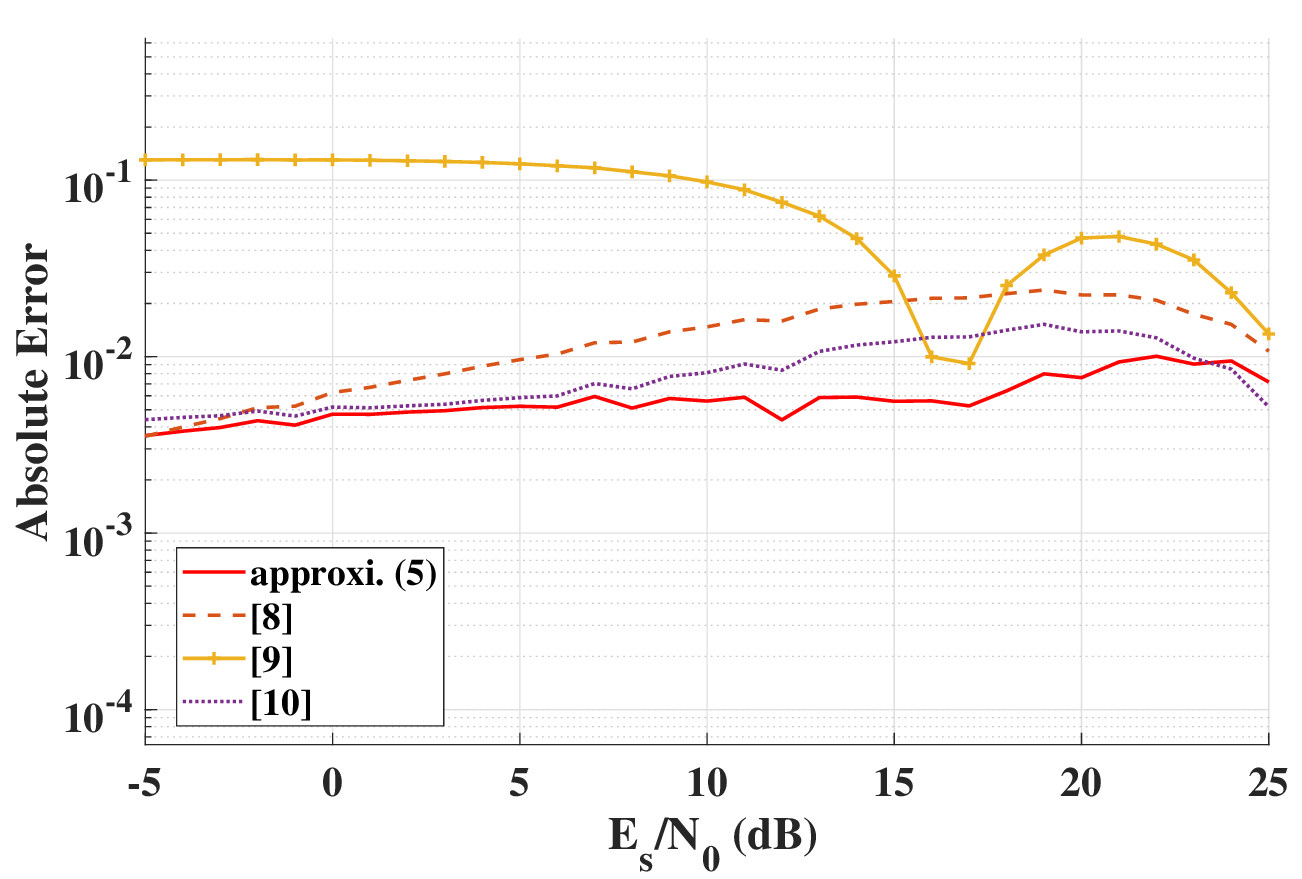}
\caption{Compares the AE v/s received SNR of 256-HQAM.}
\label{fig_6}
\end{figure}

Fig. \ref{fig_6} illustrates the AE of the exact SEP for the 256-HQAM constellation, comparing the proposed approximation with the methods presented in \cite{rugini2016symbol}, \cite{sadhwani2018simple}, and \cite{oikonomou2022error}. The AE, calculated as \( AE = {|\text{approximated SEP} - \text{exact SEP}|} \). As shown in Fig. \ref{fig_6} and Table \ref{tab_2}, the proposed approximation exhibits a lower AE than the existing methods, further validating accuracy of the results.

\begin{figure}[!t]
\centering
\includegraphics[width=8cm,height=4.5cm]{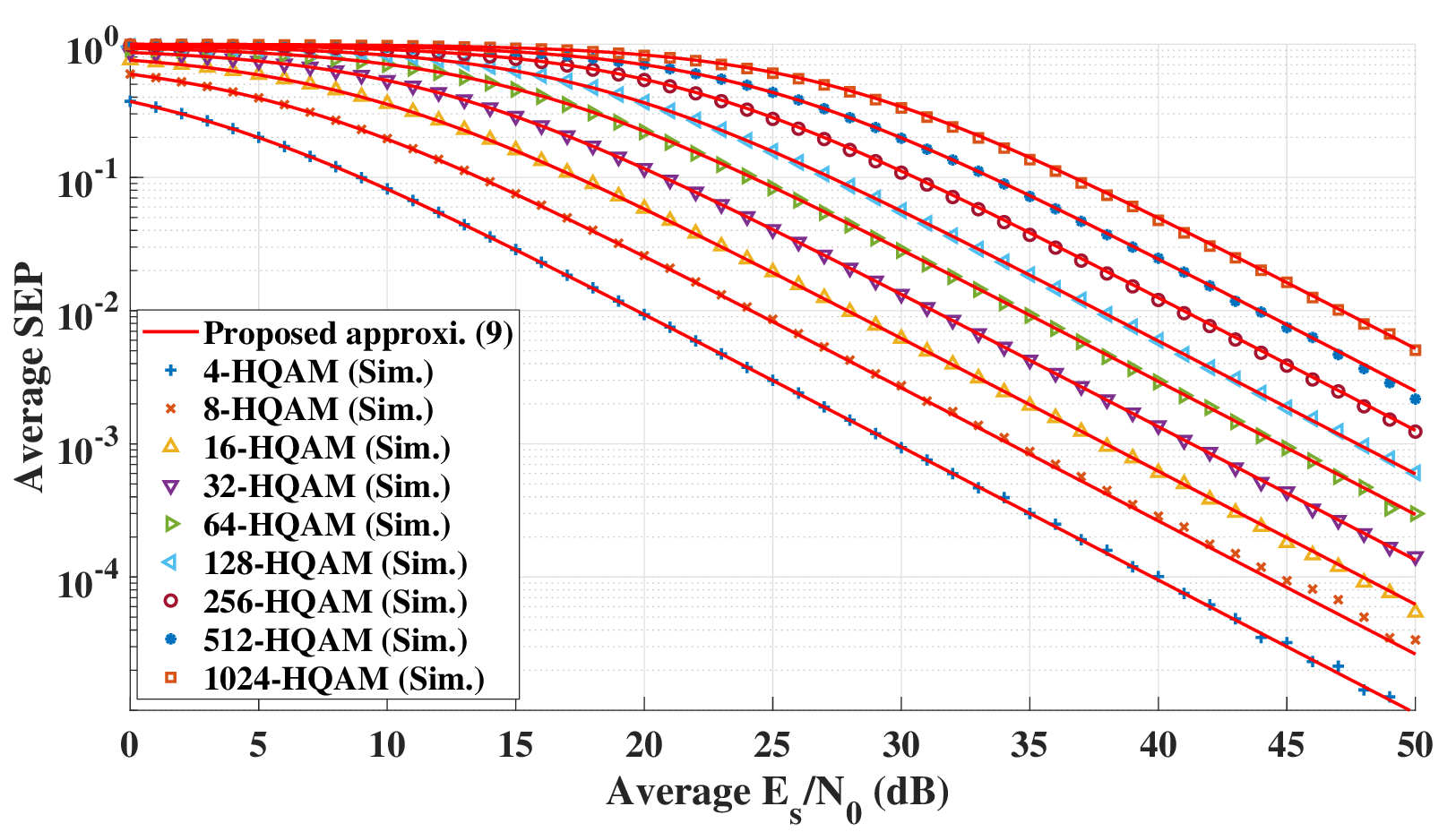}
\caption{SEP of HQAM constellation under Rayleigh fading.}
\label{fig_7}
\end{figure}

Fig. \ref{fig_7} illustrates the  SEP curves derived from (\ref{eqn_9}) for different HQAM constellations under the Rayleigh fading channel. The figure shows that the derived expressions (\ref{eqn_9}) are accurate at both low and high SNR, as demonstrated by the close alignment between the theoretical and simulated curves across all tested scenarios.

\section{Conclusion}
In this study, we derived new closed-form expressions for the SEP of general-order HQAM constellations and validated their accuracy by comparing them with simulation results in both AWGN and Rayleigh fading channels. The proposed SEP analysis is more straightforward and accurate compared to the methods in \cite{rugini2016symbol}, \cite{sadhwani2018simple}, and \cite{oikonomou2022error} and extendable to other fading scenarios. Our analytical results align closed with the simulations over a wide range of SNR values.


{\appendix[Calculation of $C$]}
From (\ref{eqn_4}), we can represent expression for $C$ as
\begin{align}
\label{eqn_10}
C=J_{1}-J_{2}+J_{3},
\end{align}
where $ J_{1}$, $ J_{2}$, and $ J_{3}$ can be obtained below as:
\begin{align}
\label{eqn_11}
  J_{1} = \frac{2}{\sqrt{2\pi}} \int_{-\infty}^{\infty}Q(\sqrt{\alpha\gamma_s})e^{-{(z-\sqrt{2\alpha\gamma_s})}^2/2}dz,
   \end{align}
Utilizing \cite[(3.323,3)]{gradshteyn2014table}, (\ref{eqn_11}) can be simplified as 
\begin{align}
   \label{eqn_12}
J_{1} = 2Q(\sqrt{\alpha\gamma_s}).
   \end{align}
\begin{align}
   \label{eqn_13}
 J_{2} = \frac{2}{\sqrt{2\pi}} \int_{-\infty}^{\infty}Q(z)e^{-{(z-\sqrt{2\alpha\gamma_s})}^2/2}dz,
   \end{align}
With the aid of \cite[(B.7)]{simon2002probability}, (\ref{eqn_13}) can be simplified as
 \begin{align}
   \label{eqn_14}
J_{2} = 2Q(\sqrt{\alpha\gamma_s}).
   \end{align}
   
\begin{equation}
   \label{eqn_15}
J_{3} = \frac{1}{\sqrt{2\pi}} \int_{-\infty}^{\infty}Q^2(z)e^{-{(z-\sqrt{2\alpha\gamma_s})}^2/2}dz,
   \end{equation}
By making use of \cite[(17)]{sofotasios2010novel} and \cite[(B.7)]{simon2002probability}, (\ref{eqn_15}) can be simplified as
 \begin{equation}
   \label{eqn_16}
J_{3} = 1.3318Q\left(\sqrt{\frac{10}{11}\alpha\gamma_s}\right)Q\left(\sqrt{\frac{1}{3}\alpha\gamma_s}\right).
   \end{equation}

\bibliography{CL}

\begin{thebibliography}{10}
\providecommand{\url}[1]{#1}
\csname url@samestyle\endcsname
\providecommand{\newblock}{\relax}
\providecommand{\bibinfo}[2]{#2}
\providecommand{\BIBentrySTDinterwordspacing}{\spaceskip=0pt\relax}
\providecommand{\BIBentryALTinterwordstretchfactor}{4}
\providecommand{\BIBentryALTinterwordspacing}{\spaceskip=\fontdimen2\font plus
\BIBentryALTinterwordstretchfactor\fontdimen3\font minus \fontdimen4\font\relax}
\providecommand{\BIBforeignlanguage}[2]{{%
\expandafter\ifx\csname l@#1\endcsname\relax
\typeout{** WARNING: IEEEtran.bst: No hyphenation pattern has been}%
\typeout{** loaded for the language `#1'. Using the pattern for}%
\typeout{** the default language instead.}%
\else
\language=\csname l@#1\endcsname
\fi
#2}}
\providecommand{\BIBdecl}{\relax}
\BIBdecl

\bibitem{singya2021survey}
P.~K. Singya, P.~Shaik, N.~Kumar, V.~Bhatia, and M.-S. Alouini, ``{A survey on higher-order QAM constellations: Technical challenges, recent advances, and future trends},'' \emph{IEEE Open Journal of the Communications Society}, vol.~2, pp. 617--655, 2021.

\bibitem{conway2013sphere}
J.~H. Conway and N.~J.~A. Sloane, \emph{{Sphere packings, lattices and groups}}.\hskip 1em plus 0.5em minus 0.4em\relax Springer Science \& Business Media, 2013, vol. 290.

\bibitem{gallager1984efficient}
R.~Gallager, G.~Lang, F.~Longstaff, and S.~Qureshi, ``{Efficient Modulation for Band-Limited Channels},'' \emph{IEEE Journal on Selected Areas in Communications}, vol.~2, no.~5, pp. 632--647, 1984.

\bibitem{park2007triangular}
S.-J. Park, ``{Triangular quadrature amplitude modulation},'' \emph{IEEE Communications Letters}, vol.~11, no.~4, pp. 292--294, 2007.

\bibitem{simon1973hexagonal}
M.~Simon and J.~Smith, ``{Hexagonal multiple phase-and-amplitude-shift-keyed signal sets},'' \emph{IEEE Transactions on Communications}, vol.~21, no.~10, pp. 1108--1115, 1973.

\bibitem{park2012performance}
S.-J. Park, ``{Performance analysis of triangular quadrature amplitude modulation in AWGN channel},'' \emph{IEEE Communications Letters}, vol.~16, no.~6, pp. 765--768, 2012.

\bibitem{qureshi2016sep}
F.~H. Qureshi, S.~A. Sheikh, Q.~U. Khan, and F.~M. Malik, ``{SEP performance of triangular QAM with MRC spatial diversity over fading channels},'' \emph{EURASIP Journal on Wireless Communications and Networking}, vol. 2016, pp. 1--16, 2016.

\bibitem{rugini2016symbol}
L.~Rugini, ``{Symbol error probability of hexagonal QAM},'' \emph{IEEE Communications Letters}, vol.~20, no.~8, pp. 1523--1526, 2016.

\bibitem{sadhwani2018simple}
D.~Sadhwani, R.~N. Yadav, S.~Aggarwal, and D.~K. Raghuvanshi, ``{Simple and accurate SEP approximation of hexagonal-QAM in AWGN channel and its application in parametric $\alpha$-$\mu$, $\eta$-$\mu$, $\lambda$-$\mu$ fading, and log-normal shadowing.}'' \emph{IET Communications (Wiley-Blackwell)}, vol.~12, no.~12, 2018.

\bibitem{oikonomou2022error}
T.~K. Oikonomou, S.~A. Tegos, D.~Tyrovolas, P.~D. Diamantoulakis, and G.~K. Karagiannidis, ``{On the error analysis of hexagonal-QAM constellations},'' \emph{IEEE Communications Letters}, vol.~26, no.~8, pp. 1764--1768, 2022.

\bibitem{proakis2008digital}
J.~G. Proakis and M.~Salehi, \emph{{Digital Communications}}.\hskip 1em plus 0.5em minus 0.4em\relax McGraw-hill, 2008.

\bibitem{beaulieu2006useful}
N.~C. Beaulieu, ``{A useful integral for wireless communication theory and its application to rectangular signaling constellation error rates},'' \emph{IEEE Transactions on Communications}, vol.~54, no.~5, pp. 802--805, 2006.

\bibitem{gradshteyn2014table}
I.~S. Gradshteyn and I.~M. Ryzhik, \emph{Table of integrals, series, and products}.\hskip 1em plus 0.5em minus 0.4em\relax Academic press, 2014.

\bibitem{simon2002probability}
M.~K. Simon, \emph{Probability distributions involving Gaussian random variables: A handbook for engineers and scientists}.\hskip 1em plus 0.5em minus 0.4em\relax Springer, 2002.

\bibitem{sofotasios2010novel}
P.~C. Sofotasios and S.~Freear, ``{Novel expressions for the Marcum and one dimensional Q-functions},'' in \emph{7th International Symposium on Wireless Communication Systems}.\hskip 1em plus 0.5em minus 0.4em\relax IEEE, 2010, pp. 736--740.

\end{thebibliography}

\bibliographystyle{IEEEtran}
\end{document}